\documentclass[conference]{IEEEtran}
\IEEEoverridecommandlockouts

\usepackage{cite}
\usepackage{amsmath,amssymb,amsfonts}
\usepackage{algorithmic}
\usepackage{graphicx}
\usepackage{textcomp}
\usepackage{xcolor}

\usepackage{physics}
\usepackage{wrapfig}
\usepackage{float}
\usepackage{tikz}
\usepackage{array}
\usepackage{qcircuit}
\usepackage{comment}
\usepackage{enumitem}
\usepackage{bm}
\usepackage{tabularx}
\usepackage{multirow}
\usepackage{soul}
\usepackage{orcidlink}
\usepackage{subfig}
\usepackage{makecell}

\begin{document}
\author{\IEEEauthorblockN{Francesco Turro \orcidlink{0000-0002-1107-2873}}
\IEEEauthorblockA{\textit{Quantum Computing Solutions}\\
\textit{Leonardo S.p.A.}\\
Via R. Pieragostini 80, Genova, Italy\\
francesco.turro@leonardo.com  }
\and
\IEEEauthorblockN{Daniele Dragoni \orcidlink{https://orcid.org/0000-0002-1644-5675}}
\IEEEauthorblockA{\textit{Quantum Computing Solutions} \\
\textit{Leonardo Hypercomputing continuum } \\
\textit{Leonardo S.p.A.}\\
Via R. Pieragostini 80, Genova, Italy } }

\title{A Resource-Efficient Quantum Framework for Graph Coloring and Chromatic Number Estimation}

\maketitle

\begin{abstract}

Many industrial optimization tasks can be modeled as graph coloring, where adjacent vertices must have different colors. This NP-hard problem is challenging for large graphs. We present a quantum encoding requiring qubits that scale logarithmically with the number of colors and linearly with vertices. Using adiabatic evolution with a novel mixer Hamiltonian and vertex terms, we compute the chromatic number and demonstrate robustness by solving constrained truck loading problems.
\end{abstract}


\section{Introduction}

Graph coloring is a fundamental problem in computational science, where the $N_G$ vertices of a graph $G$ need to be colored with a set of $\eta$ colors, such that no two adjacent vertices share the same color. This problem has a wide range of industrial applications, including scheduling~\cite{Burke2002recentresearch,Carter1008PractiteandTheory}, routing~\cite{robertazzi2000computer,Zeng_2016}, and frequency network optimization~\cite{Hale1980Frequencyassignement,aardal2007models}, all of which can be modeled as graph coloring problems. 

Illustrative industrial applications arise naturally in logistics. A representative example in logistics is the truck loading problem. Consider 
$N_G$ customer orders that must be assigned to 
$\eta$ trucks. Certain pairs of orders are incompatible, for instance, due to hazardous materials that cannot be transported together, or because different orders must be loaded at some specific time window. In addition, each truck has a limited weight capacity 
$W_{max}$.  Another relevant example arises in hospital operating room scheduling. Suppose 
$N_G$ patients require surgical procedures within specific time windows, and surgeries must be assigned to operating rooms and surgical teams. Furthermore, the time of using the operating rooms cannot pass the total daily availability time of the rooms (for example, 24h). Two surgeries that require the same room or the same team cannot overlap. This scenario can likewise be formulated as a graph coloring problem in which each vertex represents a surgery and each color corresponds to an operating room. 

In industrial applications, optimizing resource allocation and minimizing costs are critical objectives. Therefore, finding the optimal solution of the logistic problem, the smallest set of colors (i.e. trucks) that satisfies the graph coloring constraints, is of particular interest. This number is the so-called \textit{chromatic number} of the graph and is denoted by $\Gamma$ in this work. Furthermore, in real-world logistic scenarios, additional constraints may need to be incorporated into the problem, such as limitations on the truck volume space, payload, fuel capacity of each truck or the order sequences of the deliveries. These added factors significantly increase the complexity of the problem, making it more challenging and relevant to practical industrial applications.

\begin{figure*}
    \centering
\includegraphics[width=0.8\linewidth]{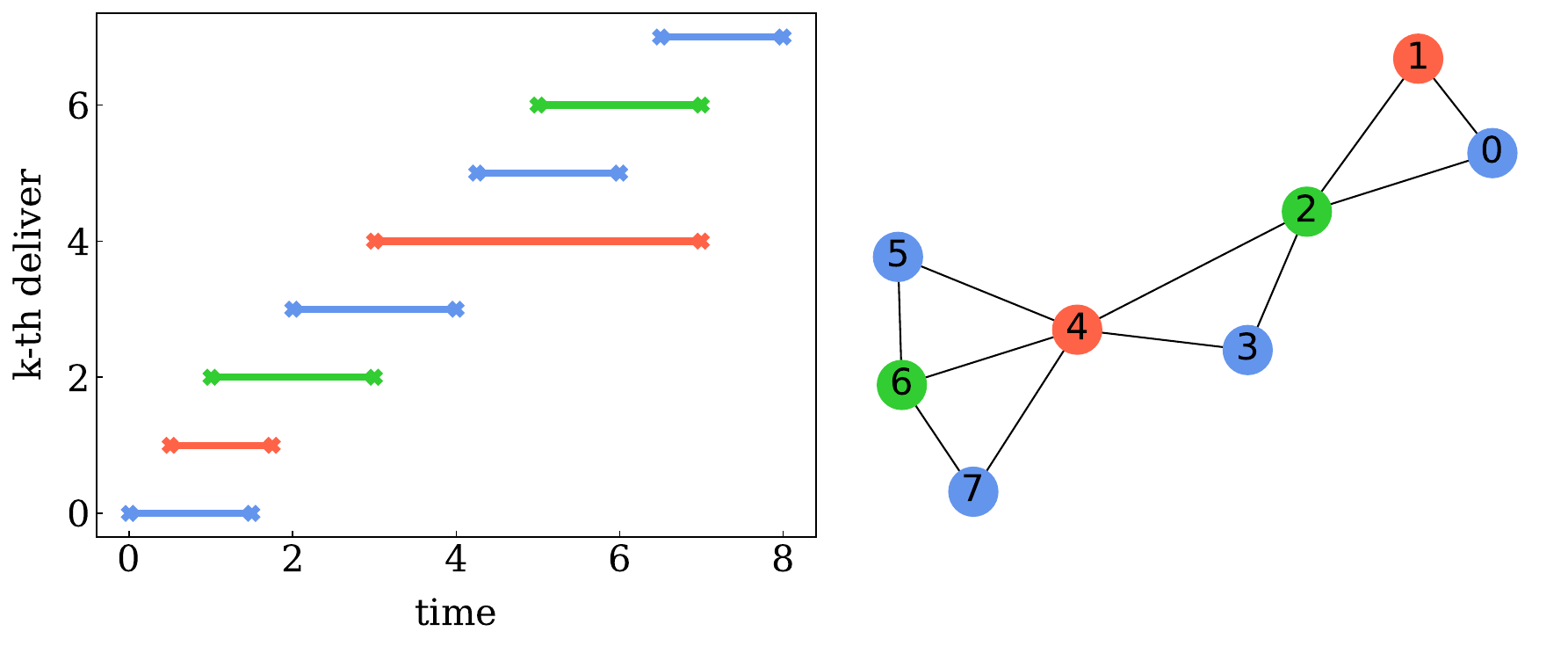}  
\caption{Illustrative example of the mapping to a graph coloring problem. 
Eight patients require surgery within the time windows shown in the left panel. 
Overlapping time windows induce conflicts, which are represented as edges in the graph shown in the right panel. The colors indicate the solution, showing that three surgical rooms are required in this instance.}
\label{fig:delivery_example}
\end{figure*}

The standard graph coloring problem is known to be NP-hard for general graphs~\cite{Garey1976somesimplifiedNP}, which means that, especially for large or complex instances, deterministic classical algorithms such as backtracking methods~\cite{Welsh1967Anupperbound,Kubale1985agenralizedimplicit} that finds the optimal solutions become computationally infeasible. Heuristic-based method~\cite{lewis2021guide}, such as the Recursive largest first algorithm~\cite{leighton1979graph} and DSatur algorithm~\cite{Brelaz1979newmethods}, do not guarantee optimal solutions, and their performance strongly depends on the structure of the graph~\cite{jensen2011graph,even1975complexity}. In recent years, quantum computing has emerged as a promising alternative approach to tackle hard combinatorial problems~\cite{abbas2024challenges}. In fact, quantum computers may offer the promise of dramatically improving the speed and efficiency of solving optimization problems, due to the possibility of exploiting inherent parallelization and memory capabilities, since all $\eta$ possible combinations  can be encoded in $\log_2(\eta)$ qubits. 

In the quantum computing framework, combinatorial optimization tasks are often conveniently formulated in the form of quadratic unconstrained binary optimization (QUBO) problems~\cite{kochenberger2014unconstrained,glover2019tutorialformulatingusingqubo}, in which the optimal solution is found minimizing the following cost function
\begin{equation}
Q(x)=\frac{1}{2} \sum_{i \neq j}  Q_{ij} x_{i} x_{j}+\sum_{i}  Q_{ii} x_{i} \,\label{eq:QUBO_simply_graph_coloring}
\end{equation}
where the entries of the real symmetric matrix $Q$ encode the optimization problem and $x_i$ are the set of binary variables. The graph coloring problem is one of the problems that can be formulated in QUBO form by introducing the binary variable $x_{i,k}$, whose value is 1 if the node $i$ is assigned the color $k$ and 0 otherwise. The graph coloring conditions are expressed through the following two constraints:
 \begin{equation}
    \sum_{k=1}^{\eta} x_{ik} x_{jk}=0 \quad \forall <i,j> \in G \,,\label{eq:constraint_color}
\end{equation}
which ensures that adjacent vertices, denoted as $<i,j>$, have different colors, and 
 \begin{equation}
\label{eq:constraint_vertex}
\sum_{k=1}^{\eta} x_{ik}=1\quad \forall i \in G\,,
 \end{equation}
 which guarantees that each vertex has only one color. Hence, the QUBO formula is given by
\begin{equation}
    Q_{gc}=\lambda_A \sum_{<i,j>\in G} \sum_{k,l=1}^\eta x_{ik} x_{jk} + \lambda_B \sum_{i\in G} (1-\sum_{k=1}^{\eta} x_{ik})^2 \,,\label{eq:QUBO_Graphcoloring}
\end{equation}
where $\lambda_{A,B}$ are Lagrangian multipliers whose ratio must be chosen correctly in order to ensure feasibility of the solution.  

The QUBO cost function can be reformulated in the form of an Ising Hamiltonian $H$, using the following map between a binary variable $x_{ik}$ and a specific qubit ($l$),
\begin{equation}
  x_{ik}=\frac{1-\sigma_{l}^z}{2}\,. \label{eq:binarymapping}
\end{equation}
where $\sigma_{l}^z$ indicates the $z-$pauli operator on the $l$-th qubit. The Ground State (GS) of resulting Ising Hamiltonian  corresponds to the optimal solution of the QUBO problem. The GS can be obtained with quantum computers implementing the Quantum Adiabatic Algorithm (QAA)~\cite{Kadowaki1998Quantumannealing,Farhi2001QAA} and the Quantum Approximate Optimization Algorithm (QAOA)~\cite{farhi2014quantumapproximateoptimizationalgorithm,blekos2024review}. The former method relies on adiabatic evolution, beginning with the known GS of a simple initial Hamiltonian, called mixer, and gradually transforming the system so that the final Hamiltonian is $H$.  The latter method is, broadly speaking, a parametrized variational version of the adiabatic evolution, with a fixed number of evolution layers, whose parameters are optimized to obtain the best approximation of the GS of $H$.

Different works~\cite{kwok2020graphcoloringquantumannealing,angkhanawin2025graphcoloringquantumoptimization,Bravyi2022hybridquantum,silva2020mapping,lodewijks2020mappingnphardnpcompleteoptimisation,liu2025efficienthybridvariationalquantum} solve the QUBO graph coloring on quantum computers (from the formulation of Eq.~\eqref{eq:QUBO_Graphcoloring}). However, this formulation is demanding in the required number of qubits, $\eta N_G$ qubits, making it computationally expensive to solve large or complex graph coloring problems with many vertices and colors. Furthermore, it can also be computationally challenging, as the same quantum circuits must be run multiple times with varying Lagrangian multipliers until a feasible solution is found, because it is not guaranteed that the found solution satisfies the graph coloring constraints with a certain Lagrangian multiplier. Other quantum methods beyond the QUBO formulation have been proposed in recent years to solve the graph coloring through quantum computers, such as the implementation of Grover’s search quantum algorithm~\cite{gaspers2023quantumalgorithmsgraphcoloring,shimizu2022exponential}, a memory efficient encoding ~\cite{Tabi2020quantumoptimizaiton,Jansen2024quditinspired,bottrill2023exploringpotentialqutritsquantum} or other mathematical formulations~\cite{tarquini2024testing,tarquini2026dronedeliverypackingproblem}.

This work follows the memory efficient approach, in which $\eta$ colors are mapped into the different states of 
$\log_2(\eta)$ qubits (or to a generic qudit with 
$\eta$ states). This encoding strictly enforces the constraint that each vertex must be assigned exactly one color, a key challenge in graph coloring.  Next, we define the interaction Hamiltonian between two adjacent vertices such that it penalizes states where the vertices share the same color.
Due to this construction, the ground state of the Hamiltonian satisfies the essential graph coloring requirement that connected vertices must have different colors.

Therefore, similar to the QUBO formulation, we then compute the ground state of this Hamiltonian applying the QAA or QAOA methods.  Past works~\cite{Tabi2020quantumoptimizaiton,Jansen2024quditinspired} suggest the use of the standard mixer $\sigma_x$.
According to Brooks' theorem~\cite{Brooks_1941}, the chromatic number satisfies $\Gamma \leq \Delta + 1$ in the worst-case scenario\footnote{This occurs for complete graphs and odd cycles.}, where $\Delta$ denotes the maximum degree of the graph. Consequently, we can restrict our simulations to $\Delta + 1$ colors. Suppose we encode the $\Delta + 1$ possible colors for each vertex using $\lceil \log_2(\Delta + 1) \rceil$ qubits, where $\lceil \cdot \rceil$ denotes the ceiling function. If $\Delta + 1$ is not a power of $2$, this encoding introduces additional computational states corresponding to colors. As a result, the use of the $\sigma_x$ mixer reduces the performance of QAA or QAOA methods because indiscriminately mixes $\Delta+1$ colors and optional colors and it gives more sub-optimal solutions (solutions with more unnecessary colors). In this work, we propose a novel mixer that operates exclusively within the feasible Hilbert space, thereby significantly improving the performance of both QAA and QAOA.

Furthermore, this formulation does not address the problem of finding the chromatic number, named $\Gamma$, which is important in many industrial applications. Indeed, if we use 
$\eta >\Gamma$ colors, the ground state of this formulation is degenerate, including all solutions of the graph coloring having $\Gamma, \Gamma+1, ...,\eta$ colors\footnote{The number of solutions with more colors is greater than those with fewer colors}. This work proposes modifying the interaction Hamiltonian by adding a simple single-vertex term that penalizes configurations with many colors. This modification ensures that the ground states of the new Hamiltonian correspond to solutions using exactly  $\Gamma$ colors, effectively solving for the chromatic number.

This paper is organized as follows: Sec.\ref{sec:graph_coloring} introduces the memory-efficient encoding of graph coloring and discusses the new mixer formulation. Sec.\ref{sec:chromatic} presents the formulation of the cost Hamiltonian designed to evaluate the chromatic number, along with numerical tests that validate the proposed method. Connecting to the truck loading problems with limited capacity mentioned in the introduction, Sec.\ref{sec:constraints} demonstrates applications with additional constraints on the graph coloring. Finally, Sec.\ref{sec:conclusion} concludes the paper.

\section{Efficient formulation of graph coloring for quantum computers }
\label{sec:graph_coloring}

We start discussing the cost Hamiltonian and mixer for the graph coloring with three colors. The generalization for generic $\eta$ colors is straightforward.

 We have to label a graph $G$ with three colors (r,g,b). We can represent these colors using the states of two qubits as $\ket{00}=r$, $\ket{01}=g$, $\ket{10}=b$; the remaining state $\ket{11}$ is considered unfeasible. This map ensures the constraint of graph coloring Eq.~\eqref{eq:constraint_vertex}. The other constraint (See Eq.~\eqref{eq:constraint_color}) can be enforced defining the interaction Hamiltonian between adjacent vertices such that it has higher energy when the two vertices have the same color or the states are unfeasible (no color). Therefore, the GS of the cost Hamiltonian gives a correct graph coloring solution (it has the minimum energy, i.e., all adjacent vertices have different colors). We define the following cost Hamiltonian $H_c(i,j)$ that acts on $i$ and $j$ vertices that are connected:
\begin{equation}
\begin{split}
H_c(i,j)= {\rm Diag}[b_1,0,0,&a_1,0,b_2,0,a_2,0,0,b_3,\\
&  a_3,a_4,a_5,a_6,a_7] \,,  \\
\end{split}
\end{equation}
where $\rm{Diag}$ denotes a diagonal matrix. The entries containing the parameters $b_i$ act on states where vertices are assigned the same color. These parameters must be positive such that they penalizes this color configuration. Conversely, the entries with parameters $a_j$ correspond to infeasible states, non color states. These values should satisfy $a_i \geq \max[b_i]$  in order to assign the highest energy penalties to invalid configurations. However, to simplify quantum circuit compilation, the values of $a_i$ and $b_i$ can be freely chosen. In this work, we set $a_i=b_j=1$.

This formulation can be generalized to the case of $\eta$ colors encoded in $n$ qubits, where \(n =\lceil \log_2(\eta)\rceil\). For any pair of states representing different colors, the corresponding matrix element is set to zero; otherwise, it is assigned a positive value. 
The final cost Hamiltonian for the $G$ graph coloring problem is given by summing all two-vertex interaction $H_c(i,j)$ for all adjacent $i$ and $j$ vertices, i.e.
\begin{equation}
    H_c=\sum_{<i,j>} H_c(i,j)\,.
\end{equation}

The ground state (GS) of the full system can be found using the QAA method. Prior works~\cite{Tabi2020quantumoptimizaiton,Jansen2024quditinspired} suggest using the Pauli $-X$ operator as the mixer. However, this standard choice reduces the efficiency of graph coloring when $\eta < 2^n$, since the mixer and initial state can generate superpositions of both feasible (valid colorings) and infeasible states. For instance, a uniform superposition over all bit strings includes many states that do not correspond to valid color assignments.

To address this issue, we propose an alternative mixer Hamiltonian with a block-diagonal structure. One block acts solely within the feasible subspace of the Hilbert space, while the other operates within the subspace of infeasible states. Our proposed mixer Hamiltonian, for the three-color case, is given by

\begin{equation}
H_m^{3,4}=QFT_{3,4}
\begin{pmatrix}
    -1 & 0 & 0 &0\\
    0 & 1 & 0 &0\\
    0 & 0 & 1 &0\\
    0 & 0 & 0 &1\\
\end{pmatrix}
QFT_{3,4}^\dagger\,, \label{eq:mixer_3color}
\end{equation}
where the central matrix indicates the eigenvalues of $H_m^{3,4}$ and the $QFT_{3,4}$ matrix indicates the eigenstate matrix given by
\begin{equation}
QFT_{3,4}=\frac{1}{\sqrt{3}}\begin{pmatrix}
    1 & 1 & 1 &0\\
    1 & \omega & \omega^2 &0\\
    1 & \omega^2 & \omega^4 &0\\
    0 & 0 & 0 &\sqrt{3}\\
\end{pmatrix}\,,
\end{equation}
with $\omega=e^{i\frac{2\pi}{3}}$. We can observe that the GS of the proposed $H_m^{3,4}$ Hamiltonian\footnote{given by the first column of the matrix $QFT_{3,4}$.} is given by the state with equal superposition of feasible states, $\ket{GS}= \frac{1}{\sqrt{3}} \left(\ket{00}+\ket{01}+\ket{10}\right)$ and this mixer does not mix the unfeasible state $\ket{11}$ with the others. 

This mixer Hamiltonian can be generalized for the case of $k$ colors encoded in $n$ qubits, as well, as
 it follows 
 \begin{equation}
    H_m^{k,n}=QFT_{k,n} D_k QFT_{k,n}^\dagger \,,\label{eq:mixer_general}
\end{equation}
where $QFT_{k,n}$ indicates a block diagonal matrix in which the first block is a $k \times k$ Quantum Fourier matrix structure and the second is the $(n-k)\times (n-k)$ identity matrix. The $D_k$ matrix indicates a diagonal matrix with entries $-1$ for the initial entry and $1$ for the others.

Moreover, the GS of $H_m^{k,n}$ can prepared on quantum processors implementing the $QFT_{k,n}$ operator. Until eight colors, we can find an efficient compilation of the $QFT_{k,n}$ and $e^{-it H_m^{k,n}}$ operators in quantum gates applying \textit{decompose} function of qiskit~\cite{qiskit2024}, that decomposes a generic 3-qubit gate in native CNOT and single qubit gates. 

We can also estimate the number of gates required for the cost Hamiltonian. Let $N_{conn}$ be the number of connected vertex pairs in the graph $G$. The real-time evolution of the cost Hamiltonian can be implemented using at most 
$N_{conn} 2^{2n}$ CNOT gates in an all-to-all connectivity architecture. This is because we apply $N_{conn}$ times the same quantum circuit for each real-time evolution of two-vertex interaction 
$H(i,j)$. This small gate can be implemented with at most $2^{2n}$ NOT gates implementing the Gray code~\cite{Mottonen2013gatedecomposition}, since the Hamiltonian 
$H(i,j)$ is expressed as a sum of linear combinations of 
$Z$-Pauli operators.

\section{Finding the chromatic number} \label{sec:chromatic}

Evaluating the chromatic number of a graph coloring is essential for industrial applications because it saves resources (money, time). In order to evaluate it, we slightly modify our Hamiltonian adding a one-vertex term. First of all, suppose we use $n$ qubits for encoding $\eta$ colors, we define the color number operator $N_i$ for the vertex $i$ of the graph $G$ as
\begin{equation}
    N_{i}=Diag[0,1,...,\eta-1,\Lambda_0,...\Lambda_{2^n-\eta}]\,,
\end{equation}
where $\eta$ indicates the total number of used colors. The set ${\Lambda_i}$ denotes a collection of tunable parameters in the infeasible space. When the proposed mixer is employed, their values do not affect the adiabatic evolution. In contrast, when using the $X$ mixer, they can be chosen to be larger in order to further penalize infeasible colors.

Summing the $N_i$ operators for each vertex, the final cost function is given by
\begin{equation}
    H^{tot}_c= \sum_{<i,j>} H_C(i,j)+ \frac{1}{\eta N_G} \sum_i N_i\,. \label{eq:Hchormatic}
\end{equation}
The one body vertex term penalizes graphs with many colors and the the factor $\frac{1}{\eta N_G}$ is essential because it ensures that the constraints of graph coloring (two close vertices must be with different colors) are always satisfied. Indeed, the cost of having same colors of adjacent vertices is still greater than the upper bound of the full one-vertex Hamiltonian
\begin{equation}
    \frac{1}{\eta  N_G} \sum_{i \in G} N_i \leq  \frac{(\eta-1)  N_G }{\eta  N_G} <1\,,
\end{equation}
where $1$ in our work is the gap between having different colors and same colors.

Following this consideration, the Ground State of $H_c^{tot}$ colors the graph with the smallest number of colors that respect the graph coloring prescriptions. This number is the chromatic number. Using this formulation, we have an advantage in the performance of the adiabatic evolution because the gap between graphs with two same colors in interacting vertices and ones with different colors remain constant to $1$.

\subsection{Numerical Test}

\begin{figure*}[h]
\centering
\includegraphics[width=1.0\linewidth]{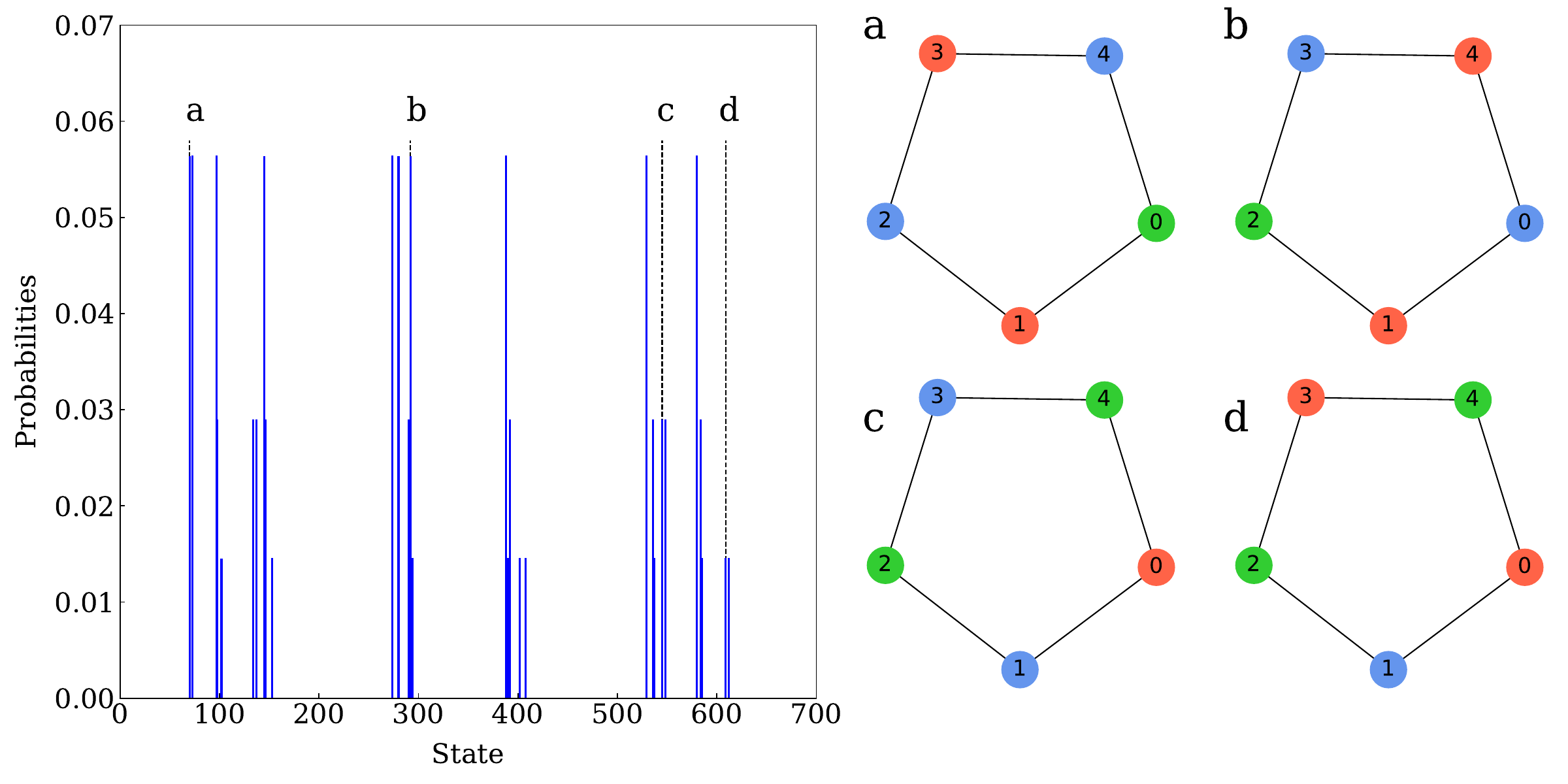}
\caption{Left panel: Sampled probabilities for a pentagon graph during adiabatic evolution. Right panels: Corresponding graph colorings for solutions labeled (a), (b), (c), and (d).}
\label{fig:adiabatic_example}
\end{figure*}

This subsection presents the results of the tests of the proposed method, starting from implementing the adiabatic evolution to evaluate the chromatic number $\Gamma$ of a pentagon graph shown in the right panels of Fig.~\ref{fig:adiabatic_example}. For this graph $\Gamma=3$, however, we use the mixer presented in Eq.~\eqref{eq:mixer_general} for $\eta=4$ colors. We consider a linear annealing schedule during the adiabatic evolution and apply a Trotter decomposition with 
$p=1000$ layers to discretize the time evolution. The total evolution time is $T_{\rm fin}=2$.

We compile the quantum circuits into gates and simulate them using the \texttt{qiskit} emulation package by IBM~\cite{qiskit2024}. The resulting probability distribution is shown in the left panel of Fig.~\ref{fig:adiabatic_example}.
The right panels show colored graphs corresponding to four sampled states, labeled (a)–(d): (a) and (b) are ground states, (c) is a first excited state, and (d) is a second excited state.

 All sampled bit strings correspond to a valid solution of the graph coloring with three colors. Interestingly, the obtained probabilities can be categorized into three distinct sets, ones with probabilities approximately $\sim 0.055$ that includes the (a) and (b) graphs, $\sim 0.03$ that includes (c), and $\sim 0.015$ that includes (d), respectively. 
 The ground states of our Hamiltonian correspond to the graphs (a) and (b). In our color encoding, where \(\ket{0} = b\), \(\ket{1} = r\), \(\ket{2} = g\), and \(\ket{3} = y\), graph (a) contains two blue, two red, and one green vertex, because the single vertex term has value as $\sim 2 \times 0 + 2 \times 1 + 1 \times 2 = 4$. 
Graph (c) has two blue, one red, and two green vertices, yielding a higher energy of $\sim 2 \times 0 + 1 \times 1 + 2 \times 2 = 5$. Similarly, graph (d) has an energy of 6 due to its color composition. We also simulated longer evolution times, in which the probabilities of sampling bit strings corresponding to cases (c) and (d) drops to zero.

\begin{figure*}[h]
\centering
\includegraphics[width=1.\linewidth]{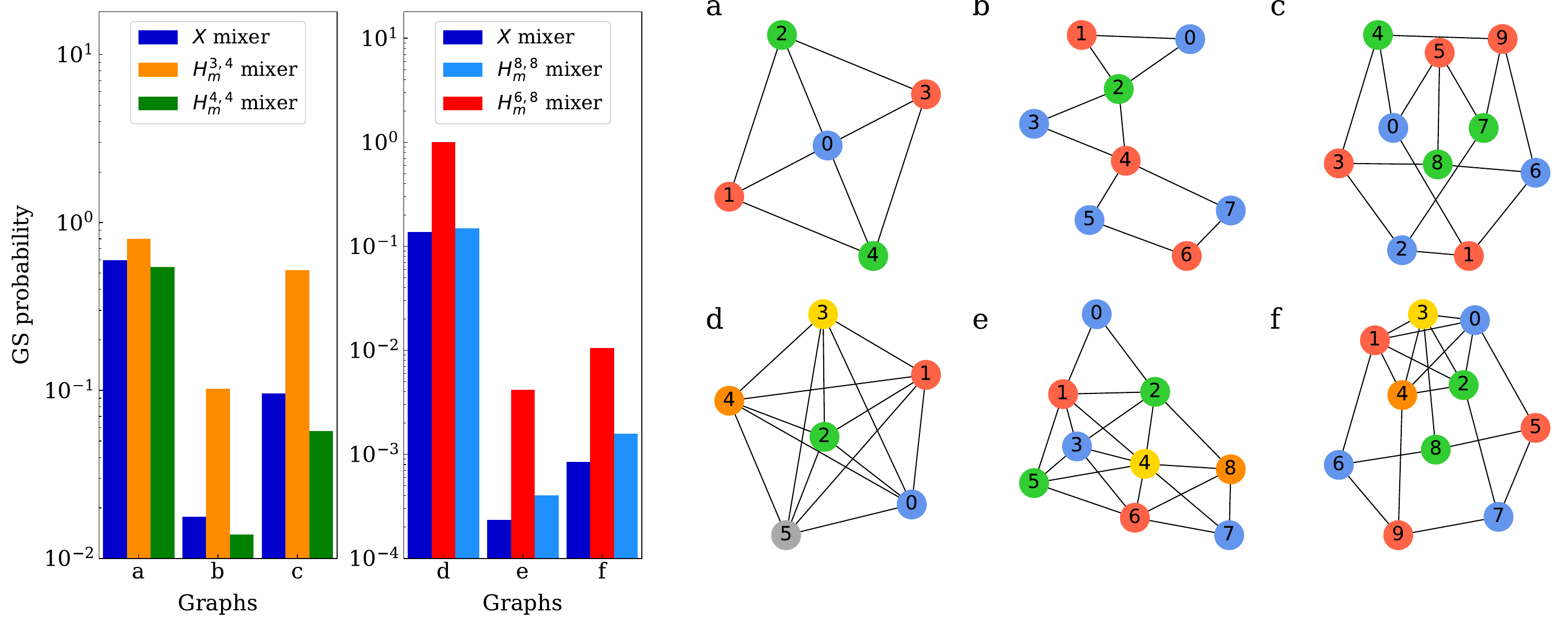}
\caption{Ground state probabilities when we use the proposed Hamiltonian to find the smallest set of colors for six graph coloring instances. We use different mixer, using the popular mixer $-X$ and the proposed mixer with a specific number of colors.
Left panel shows the obtained probabilities for different mixers. In the x-axis, letter corresponds to different graphs that are shown on the right panels. The color of the graph are given by the smallest set of colors: 3 colors for the top graphs, 6 colors for d graph and 5 colors for e and f.  }
\label{fig:chromatic_number}
\end{figure*}

We verify that the full Hamiltonian defined in Eq.~\eqref{eq:Hchormatic} correctly determines the chromatic number for different graphs. Furthermore, we test the performance of different mixer operators within the QAA framework. Specifically, we implement the QAA algorithm for the graphs shown in the right panels of Fig.~\ref{fig:chromatic_number}.

The graphs in the top panels have 
$\Gamma=3$. For these cases, we use two qubits per vertex and implement different mixers: the 
$\sigma_x$, the 
$H_m^{3,4}$ and $H_m^{4,4}$
mixers (see Eq.~\eqref{eq:mixer_general}). Panel (d) corresponds to a graph with $\Gamma=6$, while panels (e) and (f) correspond to graphs with 
$\Gamma=5$. For these graphs, we use three qubits per vertex and implement the $\sigma_x$ mixer together with the $H_m^{6,8}$ and $H_m^{8,8}$ mixers. The results are shown in Fig.~\ref{fig:chromatic_number}. We set $p=1000$ layers in the adiabatic evolution with a final time of $T_f=5$ when we use 2 qubits per colors and $T_f=10$ in the case of 3 qubits.

The left panel presents the probabilities of sampling the ground state for six graphs depicted in the right panels. Different colors of the bars correspond to the different mixers: blue for the $-X$ mixer, orange and red for the reduced mixers with 3 and 6 colors, and green and light blue for the full mixers with 4 and 8 colors. The right panels illustrate the corresponding graph colorings derived from the ground state bit strings. All identified ground states satisfy the graph coloring constraints and yield the correct chromatic number \footnote{$\Gamma=3$ for a,b,c graphs; $\Gamma=4$ for d graph; $\Gamma=5$ for e,f graphs}, as computed with the \texttt{Gurobi} solver~\cite{gurobi}.

As expected, when the graph can be colored with fewer colors than the maximum allowed by the encoding (e.g., 3 instead of 4), using a mixer acting on this smaller color set improves performance by increasing the probability of measuring the ground state. Conversely, using a mixer that spans a wider color set reduces performance but ensures that the solution satisfies the coloring constraints. Interestingly, the standard $-X$ mixer and the proposed mixer acting on the same color set yield similar probabilities of obtaining the ground state.

\section{Graph Coloring with Constraints: Truck loading problems }\label{sec:constraints}

The proposed approach only solves graph coloring problems without any constraints, for example, the truck loading application without limited capacity constraints. In fact, general use cases, like truck loading problems with capacity constraints, can be solved by adding constraints to the cost function of Eq.~\eqref{eq:Hchormatic}. Suppose that we have all trucks with the same maximum capacity $W_{max}$ and each order has a weight $w_i$.  We have to find the smallest number of trucks that can deliver the $N_G$ orders.

In the QUBO formulation, the capacity constraints can be formulated as it follows
\begin{equation}
    \sum_i w_i x_{ik} \leq W_{max},  \ \ \forall k=1,...,\eta\,.
\end{equation}

We can transform this inequality into an equality adding slack binary variables $\{y_q=0,1\}$,
\begin{equation}
    \sum_i w_i x_{ik} + \sum_{q=1}^{Y} 2^{q-1} y_q = W_{max}, \ \  \forall k=1,...,\eta\,,\label{sec:constraeq:weight_constraint_LPints}
\end{equation}
where the number of slack elements $Y$ is given by the binary representation of $W_{max}$, therefore, $\lfloor \log_2(W_{max})\rfloor+1$. Following the general mapping between QUBO formula and Ising Hamiltonian, we would encode the slack variables with extra qubits.

In this work, we did not use the QUBO formulation, so, the $x_{ik}$ binary variables must be mapped according to our encoding. This mapping can be implemented by defining the single-vertex projection for the vertex $i$ and truck $k$, $M_{i,k}$, whose elements are zero except for the 
$(k,k)$ entry, which is equal to 1.For example, $M_{i1}$, when we have four trucks, is given by
\begin{equation}
   M_{i1}= \begin{pmatrix}
        0 &0 &0 &0 \\
        0 &1 &0 &0 \\
        0 &0 &0 &0 \\
        0 &0 &0 &0 \\
    \end{pmatrix}\,.
\end{equation}

Hence, Eq.~\eqref{sec:constraeq:weight_constraint_LPints} can be transformed for the $k$th truck as 
\begin{equation}
H_P^k = \left(\sum_{i\in V} w_i M_{ik} +\sum_{q\in q_y} 2^{p-1} (1-\sigma_z^q)  - W_{max}\right)^2\,.
\end{equation}
Summing all the penalty terms for $\eta$ trucks, the final  Hamiltonian is given by
\begin{equation}
    H=H_c^{tot}+ \sum_{k=1}^{\eta} \lambda_k \left( H_P^k\right)^2\,.
\end{equation}
where $\lambda_k$ represent the Lagrangian multipliers.

In the  performed emulated tests, we avoid the use of slack variables because it decreases the number of applications that can be emulated classically, given that we can reach up to a limited number of qubits in emulation (i.e., 32 qubits). Instead, our penalty term is simply given by 
\begin{equation}
H_P^k = \left(\sum_{i\in V} w_i M_{ik} - W_{max}\right)^2\,.
\end{equation}
where we tune $\lambda_k$ in order to satisfy the constraints. The total Hamiltonian of the graph coloring with constraints is given by
\begin{equation}
        H_{\rm tot}= \sum_{<i,j>} H_C(i,j)+ \frac{\sum_i N_i}{\eta N_G}  + \sum_{k=1}^{\eta} \lambda_k H_P^k\,.\label{eq:Hamiltonian_constraints}
\end{equation}

\section{Emulation of general  truck loading problems}

\begin{table*}[t]
\begin{tabular}{|>{\centering\arraybackslash}m{4cm} 
                |>{\centering\arraybackslash}m{3cm} 
                |>{\centering\arraybackslash}m{4cm}|>{\centering\arraybackslash}m{2cm}|>{\centering\arraybackslash}m{2cm}|}
      \hline
Gurobi solution &Order weight & GS solution & $P_{c}$ with mixer $-X$ & $P_{c}$ with mixer $M(n_c,n_c)$\\
\hline
 \includegraphics[width=3.5 cm]{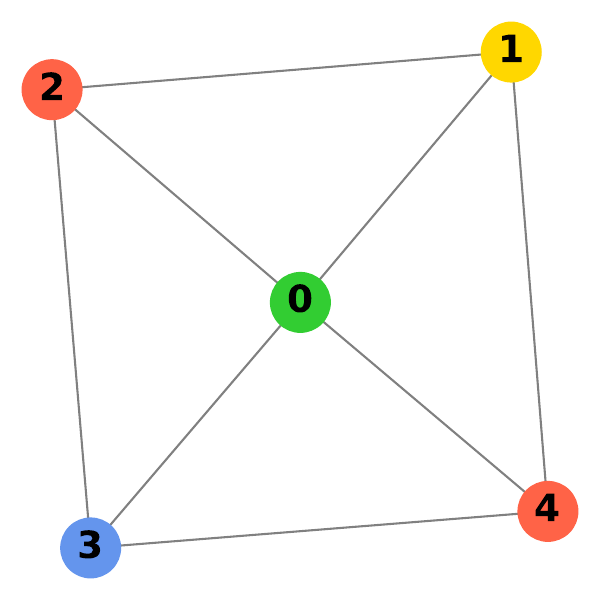} &0.44, 0.54, 0.43, 0.64, 0.37  &  \includegraphics[width=3.5 cm]{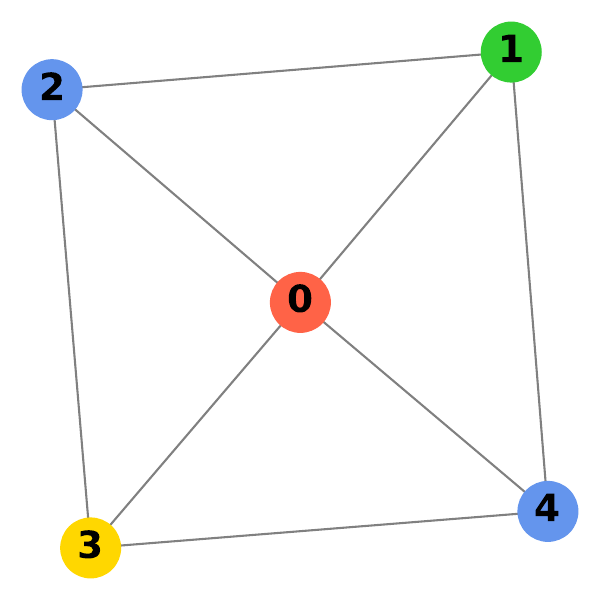}& 0.25295(6) & 0.23667(6)\\[10ex]
 
 \includegraphics[width=3.5 cm]{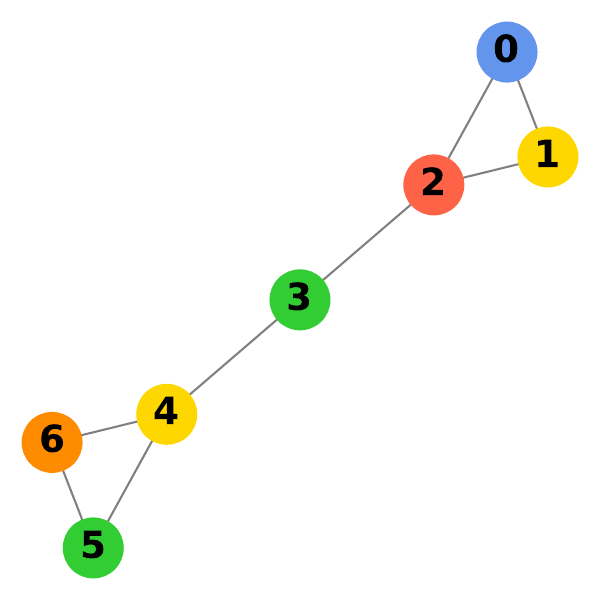} & 0.74, 0.47, 0.69, 0.4, 0.34, 0.45, 0.72 &  \includegraphics[width=3.5 cm]{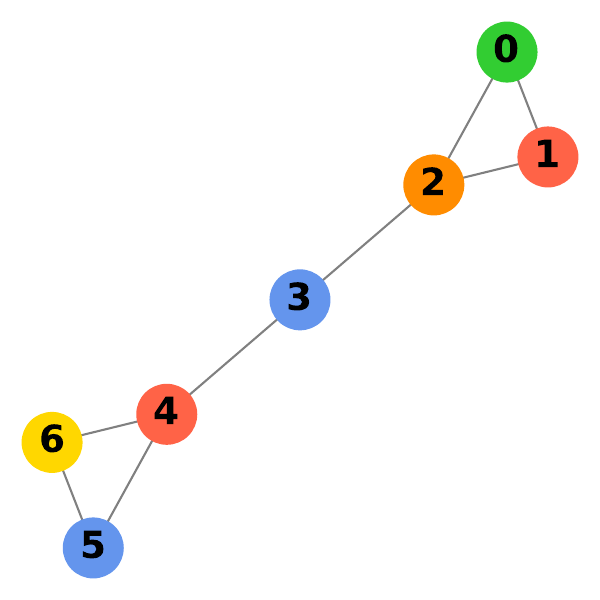}&  0.14428(5) &0.08613(4) \\[10ex]

\includegraphics[width=3.5 cm]{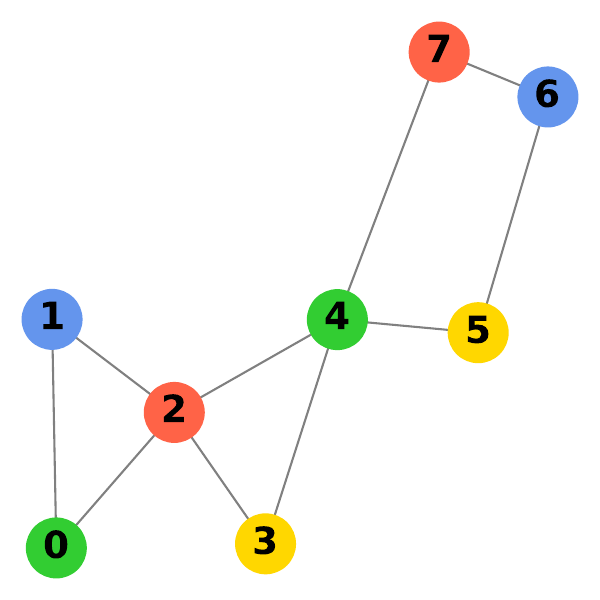} &0.41, 0.37, 0.46, 0.34, 0.42, 0.38, 0.46, 0.37 &  \includegraphics[width=3.5 cm]{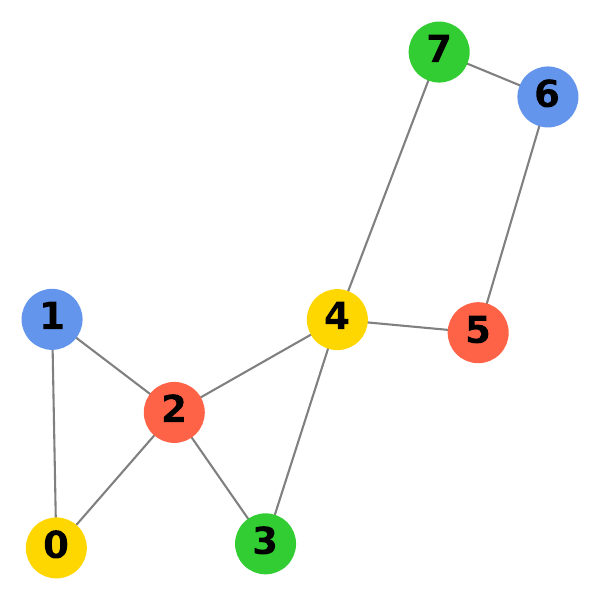}&  0.39302(6) & 0.45775(7)\\[10ex]

\includegraphics[width=3.5 cm]{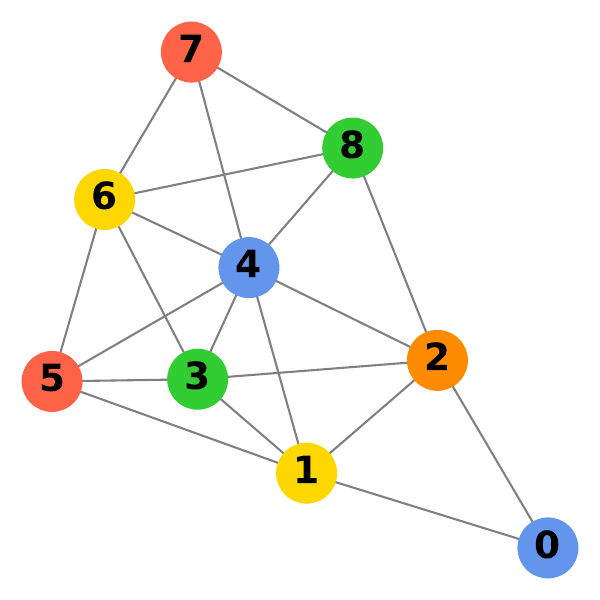} &0.35, 0.32, 0.72, 0.57, 0.63, 0.56, 0.39, 0.36, 0.42 &  \includegraphics[width=3.5cm]{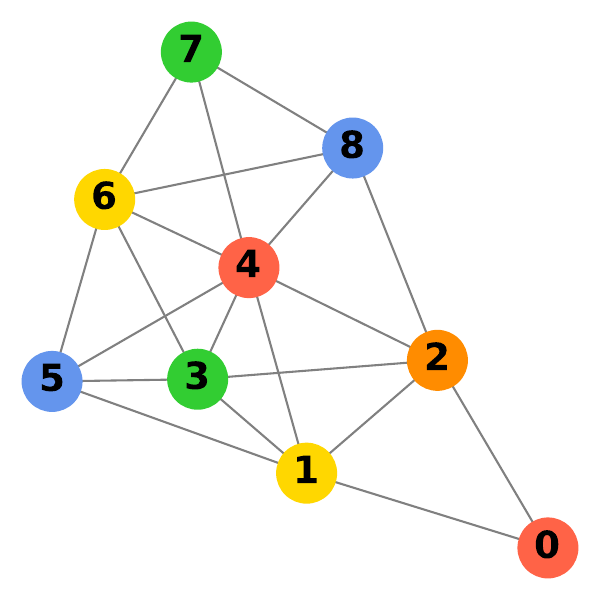}&   0.28098(6) & 0.25329(6)\\[10ex]
  \hline
\end{tabular}
\caption{Studied truck loading problems where we have to $N_G$ orders with a connectivity shown in the first/third column and each truck (color) has a limited capacity $W_{max}=1$  and each order has a specific weight reported in the second column.
The colors of first and third column are given by the solution of \texttt{gurobi} solver~\cite{gurobi} and the Ground state solution, respectively. The  fourth and fifth columns reports the obtained values of $P_c$ for two different mixers, the standard $-X$ and one of Eq.~\eqref{eq:mixer_general}. We use 8 for all graphs, except one of the first graph where we use 4 colors. In all the simulation we start from $n_c=8$ colors. The order of number in the second column indicates the weight for vertex $0$, the second to vertex $1$, and so on.}
\label{tab:results_constraints}
\end{table*}

To quantify the performance of the proposed method combined with adiabatic evolution, we define the feasible solution probability for a specific number of colors \(c\) as
\begin{equation}
    P_c = \sum_{n \in S_c} \Pr(n),
\end{equation}
where $S_c$ denotes the set of all states that use $c$ colors and satisfy the problem constraints, and $Pr(n)$ is the probability of measuring state $n$. This variable represents the total probability of obtaining all feasible solutions with exactly $c$ colors. considering all possible permutations of colors, although in our formulation the ground states (GS) correspond to a specific subset of these.

\begin{figure*}[t!]
    \centering
    \includegraphics[width=0.8\linewidth]{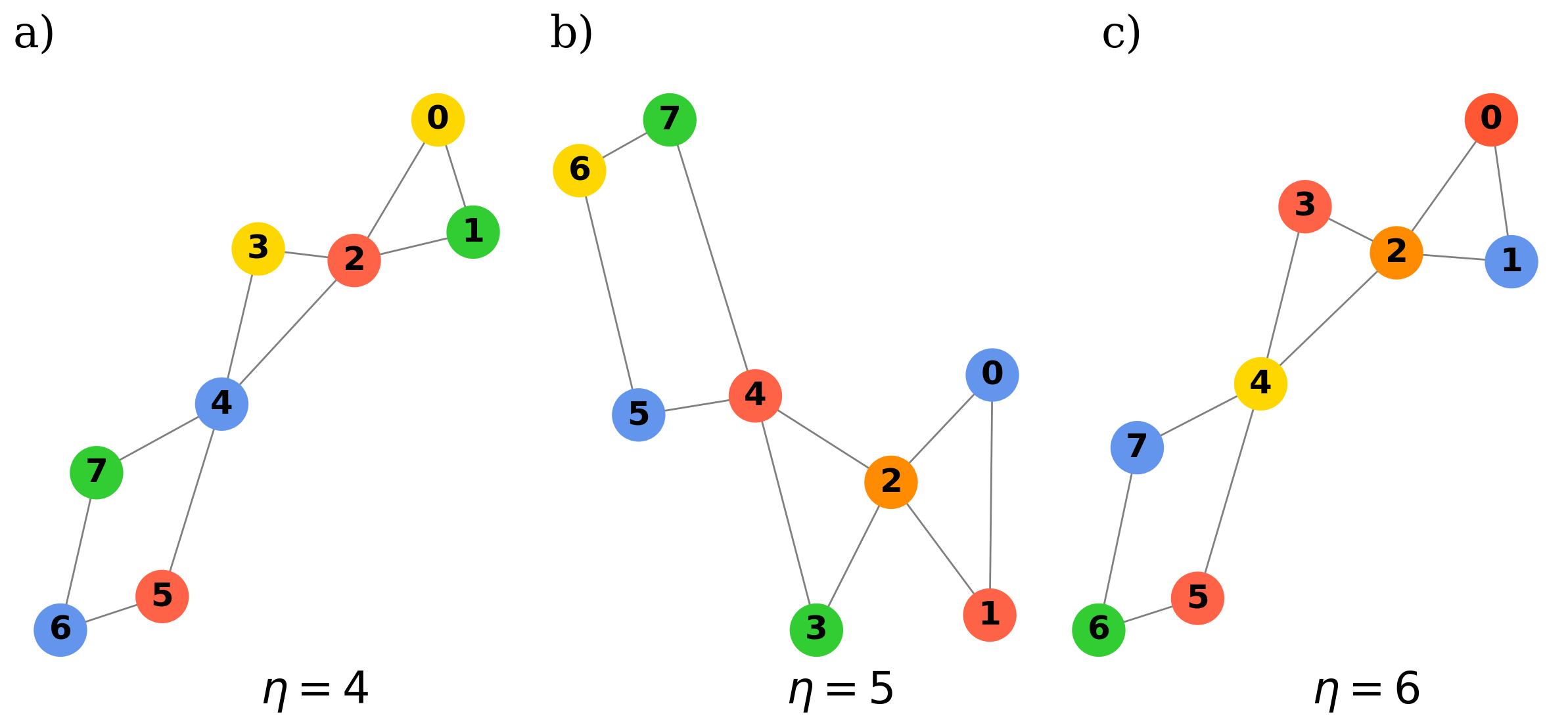}
    \caption{Ground state solutions of the same truck loading problem for different values of Lagrangian multipliers. From left to right, increasing the values causes the number of colors \(\eta\) to increase. All three graphs share the same Hilbert space dimension corresponding to 8 colors.}
    \label{fig:graphs_lambda}
\end{figure*}

We consider a truck loading problem where each trucks have a capacity of $W_{max}=1$.
 We apply the adiabatic evolution using the cost Hamiltonian defined in Eq.~\eqref{eq:Hamiltonian_constraints} to various instances. The first and second columns of Table~\ref{tab:results_constraints} reports the connectivity of the graphs and the weight for each order, respectively. The color map shown in the first column corresponds to the solution obtained from the \texttt{Gurobi} solver~\cite{gurobi}, whereas the graph of third column is derived from the GS of our Hamiltonian in Eq.~\eqref{eq:Hamiltonian_constraints}. The fourth and fifth columns reports the $P_{c=\Gamma}$ values using different mixers. Specifically, we compare the standard \(-X\) mixer with the proposed mixer from Sec.~\ref{sec:graph_coloring}. The adiabatic evolution is run with \(p=1000\) layers and final time \(T_f = 10\).

Our results indicate that the proposed method correctly finds feasible solutions for the constrained graph coloring problem, even though the GS graph coloring may differ from the \texttt{Gurobi} solution. The obtained values of $P_\Gamma$ suggest that the method is practical for industrial applications, as the correct solution can be obtained with a relatively small number of measurement shots. Furthermore, no significant differences in performance are observed between the different mixers in terms of the final probability values.

When tuning the Lagrangian multipliers associated with the capacity constraints, we observe an interesting property of our formulation with potential industrial relevance for solving constraint problems. If the multipliers are set too high, the penalty terms dominate over the one-body terms, resulting in a GS solution with a higher number of colors than the chromatic number, but it satisfies the constraints. Indeed, in energy terms, the system prefers to increase the number of colors to reduce the penalty energy. This effect is illustrated for problem of Fig.~\ref{fig:graphs_lambda} but setting different Lagrangian multipliers. Here, the ordered weights are
\begin{equation}
t_d = [0.41, 0.37, 0.46, 0.34, 0.42, 0.38, 0.46, 0.37].
\end{equation}

Panel (a) shows the optimal solution with four colors, equal to the chromatic number \(\eta=4\). Panel (b) depicts a solution with five colors, and panel (c) shows six colors. The Lagrangian multipliers used are 
\begin{equation}
\begin{split}
\lambda_a &= [1., 1., 0.5, 0.5, 0.01, 0.01, 0.01, 0.01]\\    
\lambda_b &= [1., 1., 0.5, 0.5, 0.43, 0.29, 0.15, 0.01]\,, \\
\lambda_c &= [1., 1., 1., 1., 1., 0.5, 0.5, 0.5]\,,
\end{split}
\end{equation}
for panels (a), (b), and (c), respectively. If the appropriate values of the Lagrangian multipliers are unknown, a practical approach is to start from large values and decrease them iteratively until the optimal solution is found.

Conversely, if the Lagrangian multipliers are set too small, the penalty terms become weak perturbations of the original Hamiltonian described in Sec.~\ref{sec:chromatic}, and the GS corresponds to solutions without constraints or close approximations thereof.

\section{Conclusions}
\label{sec:conclusion}

This work presents a Hamiltonian formulation of graph coloring, where we efficiently encode $\eta$ colors in $n\sim log_2\eta$ qubits. We also present a new mixer, that acts only in the feasible space, increasing significantly the performance of the adiabatic evolution, especially when we encode $\eta$ colors that are not power of 2.

Moreover, we add a specific one-vertex Hamiltonian such that the ground state is the optimal solution with the correct chromatic number, i.e., the set with the minimum number of colors. All our tests confirms the validity of the proposed Hamiltonian.

Additionally, we extend our approach to handle graph coloring problems with extra constraints, useful for logistic problems, such as the truck loading problems with limited capacity. For appropriate choices of the constraint parameters, the method yields solutions consistent with the imposed constraints. This behavior suggests that the framework may provide a viable approach for optimization problems with complex requirements.

The proposed method is memory efficiently because it grows logarithmically with the number of colors and linear with the number of vertices. The only cost is a deeper quantum circuit for implementing the one and two-body vertex-interaction terms, even though the number of CNOT gates is proportional to $\eta^2 $.

These advances represent a promising step towards scalable quantum solutions for graph coloring and connected industrial applications. With the leveraging of quantum computers in the next future, hard optimization applications can be solved with quantum computers 
With the anticipated development and deployment of quantum computers in the near future, such hard optimization problems stand to benefit from quantum advantage, unlocking new potentials beyond the capabilities of classical devices.

\section{Acknowledgment}

We thank the Quantum Computing Solution group at Leonardo S.p.A. for useful discussions.

\bibliographystyle{IEEEtran}
\bibliography{reference}

@misc{angkhanawin2025graphcoloringquantumoptimization,
      title={Graph Coloring via Quantum Optimization on a Rydberg-Qudit Atom Array}, 
      author={Toonyawat Angkhanawin and Aydin Deger and Jonathan D. Pritchard and C. Stuart Adams},
      year={2025},
      eprint={2504.08598},
      archivePrefix={arXiv},
      primaryClass={quant-ph},
      url={https://arxiv.org/abs/2504.08598}, 
}

@article{Zeng_2016,
doi = {10.1088/1751-8113/49/16/165305},
url = {https://dx.doi.org/10.1088/1751-8113/49/16/165305},
year = {2016},
month = {mar},
publisher = {IOP Publishing},
volume = {49},
number = {16},
pages = {165305},
author = {Zeng, Lishan and Zhang, Jun and Sarovar, Mohan},
title = {Schedule path optimization for adiabatic quantum computing and optimization},
journal = {Journal of Physics A: Mathematical and Theoretical}
}

@misc{kwok2020graphcoloringquantumannealing,
      title={Graph Coloring with Quantum Annealing}, 
      author={Julia Kwok and Kristen Pudenz},
      year={2020},
      eprint={2012.04470},
      archivePrefix={arXiv},
      primaryClass={quant-ph},
      url={https://arxiv.org/abs/2012.04470}, 
}

@INPROCEEDINGS{Tabi2020quantumoptimizaiton,
  author={Tabi, Zsolt and El-Safty, Kareem H. and Kallus, Zsófia and Hága, Péter and Kozsik, Tamás and Glos, Adam and Zimborás, Zoltán},
  booktitle={2020 IEEE International Conference on Quantum Computing and Engineering (QCE)}, 
  title={Quantum Optimization for the Graph Coloring Problem with Space-Efficient Embedding}, 
  year={2020},
  volume={},
  number={},
  pages={56-62},
  doi={10.1109/QCE49297.2020.00018}}

@article{Jansen2024quditinspired,
  title = {Qudit-inspired optimization for graph coloring},
  author = {Jansen, David and Heightman, Timothy and Mortimer, Luke and Perito, Ignacio and Ac\'{\i}n, Antonio},
  journal = {Phys. Rev. Appl.},
  volume = {22},
  issue = {6},
  pages = {064002},
  numpages = {12},
  year = {2024},
  month = {Dec},
  publisher = {American Physical Society},
  doi = {10.1103/PhysRevApplied.22.064002},
  url = {https://link.aps.org/doi/10.1103/PhysRevApplied.22.064002}
}

@misc{qiskit2024,
      title={Quantum computing with {Q}iskit},
      author={Javadi-Abhari, Ali and Treinish, Matthew and Krsulich, Kevin and Wood, Christopher J. and Lishman, Jake and Gacon, Julien and Martiel, Simon and Nation, Paul D. and Bishop, Lev S. and Cross, Andrew W. and Johnson, Blake R. and Gambetta, Jay M.},
      year={2024},
      doi={10.48550/arXiv.2405.08810},
      eprint={2405.08810},
      archivePrefix={arXiv},
      primaryClass={quant-ph}
}

@misc{gurobi,
  author = {{Gurobi Optimization, LLC}},
  title = {{Gurobi Optimizer Reference Manual}},
  year = 2024,
  url = "https://www.gurobi.com"
}

@ARTICLE{Hale1980Frequencyassignement,
  author={Hale, W.K.},
  journal={Proceedings of the IEEE}, 
  title={Frequency assignment: Theory and applications}, 
  year={1980},
  volume={68},
  number={12},
  pages={1497-1514},
  doi={10.1109/PROC.1980.11899}}

@article{Burke2002recentresearch,
title = {Recent research directions in automated timetabling},
journal = {European Journal of Operational Research},
volume = {140},
number = {2},
pages = {266-280},
year = {2002},
issn = {0377-2217},
doi = {https://doi.org/10.1016/S0377-2217(02)00069-3},
url = {https://www.sciencedirect.com/science/article/pii/S0377221702000693},
author = {Edmund Kieran Burke and Sanja Petrovic}
}

@InProceedings{Carter1008PractiteandTheory,
author="Carter, Michael W.
and Laporte, Gilbert",
editor="Burke, Edmund
and Carter, Michael",
title="Recent developments in practical course timetabling",
booktitle="Practice and Theory of Automated Timetabling II",
year="1998",
publisher="Springer Berlin Heidelberg",
address="Berlin, Heidelberg",
pages="3--19",
}

@book{robertazzi2000computer,
  title={Computer networks and systems: queueing theory and performance evaluation},
  author={Robertazzi, Thomas G},
  year={2000},
  publisher={Springer Science \& Business Media},
doi={https://doi.org/10.1007/978-1-4612-1164-8}
}

@article{aardal2007models,
  title={Models and solution techniques for frequency assignment problems},
  author={Aardal, Karen I and Van Hoesel, Stan PM and Koster, Arie MCA and Mannino, Carlo and Sassano, Antonio},
  journal={Annals of Operations Research},
  volume={153},
  pages={79--129},
  year={2007},
  publisher={Springer},
doi={https://doi.org/10.1007/s10479-007-0178-0}
}

@article{abbas2024challenges,
  title={Challenges and opportunities in quantum optimization},
  author={Abbas, Amira and Ambainis, Andris and Augustino, Brandon and B{\"a}rtschi, Andreas and Buhrman, Harry and Coffrin, Carleton and Cortiana, Giorgio and Dunjko, Vedran and Egger, Daniel J and Elmegreen, Bruce G and others},
  journal={Nature Reviews Physics},
  pages={1--18},
  year={2024},
  publisher={Nature Publishing Group UK London},
doi={10.1038/s42254-024-00770-9}
}

@misc{farhi2014quantumapproximateoptimizationalgorithm,
      title={A Quantum Approximate Optimization Algorithm}, 
      author={Edward Farhi and Jeffrey Goldstone and Sam Gutmann},
      year={2014},
      eprint={1411.4028},
      archivePrefix={arXiv},
      primaryClass={quant-ph},
      url={https://arxiv.org/abs/1411.4028}, 
}

@article{blekos2024review,
  title={A review on quantum approximate optimization algorithm and its variants},
  author={Blekos, Kostas and Brand, Dean and Ceschini, Andrea and Chou, Chiao-Hui and Li, Rui-Hao and Pandya, Komal and Summer, Alessandro},
  journal={Physics Reports},
  volume={1068},
  pages={1--66},
  year={2024},
  publisher={Elsevier},
doi={https://doi.org/10.1016/j.physrep.2024.03.002}
}

@article{Kadowaki1998Quantumannealing,
  title = {Quantum annealing in the transverse Ising model},
  author = {Kadowaki, Tadashi and Nishimori, Hidetoshi},
  journal = {Phys. Rev. E},
  volume = {58},
  issue = {5},
  pages = {5355--5363},
  numpages = {0},
  year = {1998},
  month = {Nov},
  publisher = {American Physical Society},
  doi = {10.1103/PhysRevE.58.5355},
  url = {https://link.aps.org/doi/10.1103/PhysRevE.58.5355}
}

@article{Bravyi2022hybridquantum,
  doi = {10.22331/q-2022-03-30-678},
  url = {https://doi.org/10.22331/q-2022-03-30-678},
  title = {Hybrid quantum-classical algorithms for approximate graph coloring},
  author = {Bravyi, Sergey and Kliesch, Alexander and Koenig, Robert and Tang, Eugene},
  journal = {{Quantum}},
  issn = {2521-327X},
  publisher = {{Verein zur F{\"{o}}rderung des Open Access Publizierens in den Quantenwissenschaften}},
  volume = {6},
  pages = {678},
  month = mar,
  year = {2022}
}

@article{silva2020mapping,
  title={Mapping graph coloring to quantum annealing},
  author={Silva, Carla and Aguiar, Ana and Lima, Priscila MV and Dutra, In{\^e}s},
  journal={Quantum Machine Intelligence},
  volume={2},
  pages={1--19},
  year={2020},
  publisher={Springer},
doi={https://doi.org/10.1007/s42484-020-00028-4}
}

@article{Garey1976somesimplifiedNP,
title = {Some simplified NP-complete graph problems},
journal = {Theoretical Computer Science},
volume = {1},
number = {3},
pages = {237-267},
year = {1976},
issn = {0304-3975},
doi = {https://doi.org/10.1016/0304-3975(76)90059-1},
url = {https://www.sciencedirect.com/science/article/pii/0304397576900591},
author = {M.R. Garey and D.S. Johnson and L. Stockmeyer},

}

@misc{glover2019tutorialformulatingusingqubo,
      title={A Tutorial on Formulating and Using QUBO Models}, 
      author={Fred Glover and Gary Kochenberger and Yu Du},
      year={2019},
      eprint={1811.11538},
      archivePrefix={arXiv},
      primaryClass={cs.DS},
      url={https://arxiv.org/abs/1811.11538}, 
}

@misc{lodewijks2020mappingnphardnpcompleteoptimisation,
      title={Mapping NP-hard and NP-complete optimisation problems to Quadratic Unconstrained Binary Optimisation problems}, 
      author={Bas Lodewijks},
      year={2020},
      eprint={1911.08043},
      archivePrefix={arXiv},
      primaryClass={cs.DS},
      url={https://arxiv.org/abs/1911.08043}, 
}

@article{kochenberger2014unconstrained,
  title={The unconstrained binary quadratic programming problem: a survey},
  author={Kochenberger, Gary and Hao, Jin-Kao and Glover, Fred and Lewis, Mark and L{\"u}, Zhipeng and Wang, Haibo and Wang, Yang},
  journal={Journal of combinatorial optimization},
  volume={28},
  pages={58--81},
doi={https://doi.org/10.1007/s10878-014-9734-0},
  year={2014},
  publisher={Springer}
}

@misc{bottrill2023exploringpotentialqutritsquantum,
      title={Exploring the Potential of Qutrits for Quantum Optimization of Graph Coloring}, 
      author={Gabriel Bottrill and Mudit Pandey and Olivia Di Matteo},
      year={2023},
      eprint={2308.08050},
      archivePrefix={arXiv},
      primaryClass={quant-ph},
      url={https://arxiv.org/abs/2308.08050}, 
}

@misc{gaspers2023quantumalgorithmsgraphcoloring,
      title={Quantum Algorithms for Graph Coloring and other Partitioning, Covering, and Packing Problems}, 
      author={Serge Gaspers and Jerry Zirui Li},
      year={2023},
      eprint={2311.08042},
      archivePrefix={arXiv},
      primaryClass={cs.DS},
      url={https://arxiv.org/abs/2311.08042}, 
}

@article{shimizu2022exponential,
  title={Exponential-time quantum algorithms for graph coloring problems},
  author={Shimizu, Kazuya and Mori, Ryuhei},
  journal={Algorithmica},
  volume={84},
  number={12},
  pages={3603--3621},
  year={2022},
  publisher={Springer},
doi={https://doi.org/10.1007/s00453-022-00976-2}
}

@misc{liu2025efficienthybridvariationalquantum,
      title={Efficient hybrid variational quantum algorithm for solving graph coloring problem}, 
      author={Dongmei Liu and Jian Li and Xiubo Cheng and Shibing Zhang and Yan Chang and Lili Yan},
      year={2025},
      eprint={2504.21335},
      archivePrefix={arXiv},
      primaryClass={quant-ph},
      url={https://arxiv.org/abs/2504.21335}, 
}

@article{Brelaz1979newmethods,
author = {Br\'{e}laz, Daniel},
title = {New methods to color the vertices of a graph},
year = {1979},
issue_date = {April 1979},
publisher = {Association for Computing Machinery},
address = {New York, NY, USA},
volume = {22},
number = {4},
issn = {0001-0782},
url = {https://doi.org/10.1145/359094.359101},
doi = {10.1145/359094.359101},
journal = {Commun. ACM},
month = apr,
pages = {251–256},
numpages = {6}
}

@article{Welsh1967Anupperbound,
    author = {Welsh, D. J. A. and Powell, M. B.},
    title = {An upper bound for the chromatic number of a graph and its application to timetabling problems},
    journal = {The Computer Journal},
    volume = {10},
    number = {1},
    pages = {85-86},
    year = {1967},
    month = {01},
    issn = {0010-4620},
    doi = {10.1093/comjnl/10.1.85},
    url = {https://doi.org/10.1093/comjnl/10.1.85},
    eprint = {https://academic.oup.com/comjnl/article-pdf/10/1/85/1069035/100085.pdf},
}

@book{jensen2011graph,
  title={Graph coloring problems},
  author={Jensen, Tommy R and Toft, Bjarne},
  year={2011},
  publisher={John Wiley \& Sons}
}

@inproceedings{even1975complexity,
  title={On the complexity of time table and multi-commodity flow problems},
  author={Even, Shimon and Itai, Alon and Shamir, Adi},
  booktitle={16th annual symposium on foundations of computer science (sfcs 1975)},
  pages={184--193},
  year={1975},
  organization={IEEE}
}

@article{
Farhi2001QAA,
author = {Edward Farhi  and Jeffrey Goldstone  and Sam Gutmann  and Joshua Lapan  and Andrew Lundgren  and Daniel Preda },
title = {A Quantum Adiabatic Evolution Algorithm Applied to Random Instances of an NP-Complete Problem},
journal = {Science},
volume = {292},
number = {5516},
pages = {472-475},
year = {2001},
doi = {10.1126/science.1057726},
URL = {https://www.science.org/doi/abs/10.1126/science.1057726},
eprint = {https://www.science.org/doi/pdf/10.1126/science.1057726}}

@article{Mottonen2013gatedecomposition,
  title = {Quantum Circuits for General Multiqubit Gates},
  author = {M\"ott\"onen, Mikko and Vartiainen, Juha J. and Bergholm, Ville and Salomaa, Martti M.},
  journal = {Phys. Rev. Lett.},
  volume = {93},
  issue = {13},
  pages = {130502},
  numpages = {4},
  year = {2004},
  month = {Sep},
  publisher = {American Physical Society},
  doi = {10.1103/PhysRevLett.93.130502},
  url = {https://link.aps.org/doi/10.1103/PhysRevLett.93.130502}
}

@article{leighton1979graph,
  title={A graph coloring algorithm for large scheduling problems},
  author={Leighton, Frank Thomson},
  journal={J Res Natl Bur Stand},
  volume={84},
  number={6},
  pages={489},
  year={1979},
  doi={https://doi.org/10.6028/jres.084.024}
}

@article{Kubale1985agenralizedimplicit,
author = {Kubale, Marek and Jackowski, Boguslaw},
title = {A generalized implicit enumeration algorithm for graph coloring},
year = {1985},
issue_date = {April 1985},
publisher = {Association for Computing Machinery},
address = {New York, NY, USA},
volume = {28},
number = {4},
issn = {0001-0782},
url = {https://doi.org/10.1145/3341.3350},
doi = {10.1145/3341.3350},
journal = {Commun. ACM},
month = apr,
pages = {412–418},
numpages = {7}
}

@book{lewis2021guide,
  title={Guide to graph colouring},
  author={Lewis, Rhyd},
  year={2021},
  publisher={Springer},
  doi={https://doi.org/10.1007/978-3-319-25730-3}
}

@article{Brooks_1941, title={On colouring the nodes of a network}, volume={37}, DOI={10.1017/S030500410002168X}, number={2}, journal={Mathematical Proceedings of the Cambridge Philosophical Society}, author={Brooks, R. L.}, year={1941}, pages={194–197}}

@article{tarquini2024testing,
  title={Testing quantum and simulated annealers on the drone delivery packing problem},
  author={Tarquini, Sara and Dragoni, Daniele and Vandelli, Matteo and Tudisco, Francesco},
  journal={Quantum Mach. Intell.},
   volume = {8},
 number = {14},
  year={2026},
  doi = {https://doi.org/10.1007/s42484-026-00362-z},
}

@misc{tarquini2026dronedeliverypackingproblem,
      title={Drone delivery packing problem on a neutral-atom quantum computer}, 
      author={Sara Tarquini and Matteo Vandelli and Francesco Ferrari and Daniele Dragoni and Francesco Tudisco},
      year={2026},
      eprint={2602.15487},
      archivePrefix={arXiv},
      primaryClass={quant-ph},
      url={https://arxiv.org/abs/2602.15487}, 
}

\appendix
\section{Data}
Table~\ref{tab:chromatic_probabilities_GS} reports the data plotted in Fig.~\ref{fig:chromatic_number}.

\begin{table*}[h]
    \centering
    \begin{tabular}{|c|c|cc|cc|}
    \hline
      Graph & mixer X & Mixer M(3,4) & Mixer M(4,4) & Mixer M(6,8) & Mixer M(8,8)\\ 
    \hline

      a   & 0.59571(7) & 0.79920(6) & 0.54206(7)& &  \\
      b   & 0.1763(4)& 0.3754(5)&0.1884(4)&& \\
      c   &0.0958(3) & 0.5208(5)& 0.0573(3)& & \\
      d   & 0.1375(3) & && 0.99977(2) & 0.1497(3)\\
      e   & 0.000236(2)& 0.02291(2)&0.004164(9)&& \\
      f & 0.000851(4) & &&0.010525(14)&0.001579(6) \\
    \hline

    \end{tabular}
    \caption{Values of Ground State probabilities plotted in Fig.~\ref{fig:chromatic_number}}
    \label{tab:chromatic_probabilities_GS}
\end{table*}

\end{document}